\documentclass[12pt]{article}
\usepackage{amsmath,epsfig,euscript}

\def\F{{\EuScript F}}

\begin{document}

\begin{titlepage}

\begin{flushright}
CLNS~05/1922\\
MIT-CTP~3668\\
{\tt hep-ph/0508178}\\[0.2cm]
August 16, 2005
\end{flushright}

\vspace{0.7cm}
\begin{center}
\Large\bf\boldmath
A two-loop relation between inclusive radiative and semileptonic $B$-decay 
spectra
\unboldmath
\end{center}

\vspace{0.8cm}
\begin{center}
{\sc Bj\"orn O. Lange$^a$, Matthias Neubert$^{b,c}$, and Gil Paz$^b$}\\
\vspace{0.4cm}
{\sl $^a$\,Center for Theoretical Physics,  
Massachusetts Institute of Technology\\
Cambridge, MA 02139, U.S.A.\\[0.3cm]
$^b$\,Institute for High-Energy Phenomenology\\
Newman Laboratory for Elementary-Particle Physics, Cornell University\\
Ithaca, NY 14853, U.S.A.\\[0.3cm]
$^c$\,Institut f\"ur Theoretische Physik, Universit\"at Heidelberg\\
Philosophenweg 16, D--69120 Heidelberg, Germany}
\end{center}

\vspace{1.0cm}
\begin{abstract}
\vspace{0.2cm}
\noindent 
A shape-function independent relation is derived between the partial 
$\bar B\to X_u\,l^-\bar\nu$ decay rate with a cut on 
$P_+=E_X-|\vec P_X|\le\Delta$ and 
a weighted integral over the normalized $\bar B\to X_s\gamma$ photon-energy 
spectrum. The leading-power contribution to the weight function is calculated 
at next-to-next-to-leading order in renormalization-group improved 
perturbation theory, including exact two-loop matching corrections at the 
scale $\mu_i\sim\sqrt{m_b\Lambda_{\rm QCD}}$. 
The overall normalization of the weight 
function is obtained up to yet unknown corrections of order $\alpha_s^2(m_b)$.
Power corrections from phase-space factors are included exactly, while the
remaining subleading contributions are included at first order in 
$\Lambda_{\rm QCD}/m_b$. At 
this level unavoidable hadronic uncertainties enter, which are estimated in a 
conservative way. The combined theoretical accuracy in the extraction of 
$|V_{ub}|$ is at the level of 5\% if a value of $\Delta$ near the charm 
threshold can be achieved experimentally.
\end{abstract}
\vfil

\end{titlepage}

\section{Introduction}

In the recent past, much progress has been made in the theoretical
understanding of inclusive charmless $B$ decays near the kinematic
endpoint of small $P_+=E_X-|\vec P_X|$, where $E_X$ and $\vec P_X$ are
the energy and momentum of the final-state hadronic system in the
$B$-meson rest frame. In $\bar B\to X_s\gamma$ decays the $P_+$
variable is related to the $B$-meson mass and the photon energy, 
$P_+=M_B-2E_\gamma$, and the measurement of its spectrum leads directly
to the extraction of the leading hadronic structure function, called
the shape function \cite{Neubert:1993ch,Neubert:1993um,Bigi:1993ex}. 
The $P_+$ spectrum in semileptonic $\bar B\to X_u\,l^-\bar\nu$ decays, on the 
other hand, enables us to determine $|V_{ub}|$ 
\cite{Mannel:1999gs,Aglietti:2002md,Bosch:2004bt}, but this requires a precise 
knowledge of the shape function. One approach for measuring $|V_{ub}|$ is to 
first extract the shape function from the $\bar B\to X_s\gamma$
photon spectrum, and then to use this information for predictions
of event distributions in $\bar B\to X_u\,l^-\bar\nu$. A comprehensive
description of this program has been presented in
\cite{Lange:2005yw}. Equivalently, it is possible to eliminate the
shape function in $\bar B\to X_u\,l^-\bar\nu$ decay rates in favor of
the $\bar B\to X_s\gamma$ photon-energy spectrum. This idea was first put
forward in \cite{Neubert:1993um} and later refined in
\cite{Leibovich:1999xf,Leibovich:2000ey,Neubert:2001sk,Hoang:2005pj}. Partial
$\bar B\to X_u\,l^-\bar\nu$ decay rates are then given as weighted
integrals over the $\bar B\to X_s\gamma$ photon-energy spectrum,
\begin{equation}\label{eq:relation}
   \Gamma_u(\Delta)
   = \underbrace{\int_0^\Delta\!dP_+\,
    \frac{d\Gamma_u}{dP_+}}_{\hbox{\footnotesize exp.\ input}}
   = |V_{ub}|^2 \int_0^\Delta\!dP_+
    \underbrace{\phantom{\frac{1}{\Gamma_s(E_*)}}\hspace{-12mm}
    W(\Delta,P_+)}_{\hbox{\footnotesize theory}}\,
    \underbrace{\frac{1}{\Gamma_s(E_*)}\,
    \frac{d\Gamma_s}{dP_+}}_{\hbox{\footnotesize exp.\ input}} \,,
\end{equation}
where the weight function $W(\Delta,P_+)$ is perturbatively calculable
at leading power in $\Lambda_{\rm QCD}/m_b$. A comparison of both
sides of the equation determines the CKM matrix element $|V_{ub}|$
directly. For the measurement of the left-hand side 
to be free of charm background, $\Delta$ must be
less than $M_D^2/M_B\approx 0.66$\,GeV. However, the $P_+$ spectrum in
$\bar B\to X_u\,l^-\bar\nu$ decays displays many of the features of
the charged-lepton energy spectrum, so that it is not inconceivable that the 
cut can be further relaxed for the same reasons that experimenters are able 
to relax the lepton cut beyond the charm threshold.
We stress that for an application of relation (\ref{eq:relation}) a 
measurement of the $\bar B\to X_s\gamma$ photon spectrum is needed only for 
$E_\gamma\ge\frac12(M_B-\Delta)\approx 2.3$\,GeV (or slightly lower, if 
the cut is relaxed into the charm region). This high-energy part of the
spectrum has already been measured with good precision.

Previous authors \cite{Neubert:1993um,Leibovich:1999xf,Leibovich:2000ey,%
Neubert:2001sk,Hoang:2005pj} have considered relations such as 
(\ref{eq:relation}) in the slightly different form
\begin{equation}\label{oldrelation}
   \underbrace{\int_0^\Delta\!dP_+\,
    \frac{d\Gamma_u}{dP_+}}_{\hbox{\footnotesize exp.\ input}}
   = \frac{|V_{ub}|^2}{|V_{tb} V_{ts}^*|^2} \int_0^\Delta\!dP_+
    \underbrace{\phantom{\frac{1}{\Gamma_s(E_*)}}\hspace{-12mm}
    \widetilde W(\Delta,P_+)}_{\hbox{\footnotesize theory}}\!
    \underbrace{\frac{d\Gamma_s}{dP_+}}_{\hbox{\footnotesize exp.\ input}}
   \hspace{-0.3cm} .
\end{equation}
Normalizing the photon spectrum by the total\footnote{Due to an
unphysical soft-photon singularity, the total decay rate is commonly
defined to include all events with photon energies above 
$E_*=m_b/20$ \cite{Kagan:1998ym}.} 
rate $\Gamma_s(E_*)$ as done in (\ref{eq:relation}) has several advantages. 
Firstly, it is a known fact that event fractions in $\bar B\to X_s\gamma$ 
decay can be calculated with better accuracy than partial decay rates (see 
\cite{Neubert:2004dd} for a recent discussion), and likewise the normalized 
rate does not suffer from the relatively large experimental error on the 
total branching ratio. Secondly, relation 
(\ref{eq:relation}) is independent of the CKM factor $|V_{tb} V_{ts}^*|$. 
Thirdly, unlike the total $\bar B\to X_s\gamma$ decay rate, the shape of the 
photon spectrum is rather insensitive to possible New Physics 
contributions \cite{Kagan:1998ym}, which could distort the outcome of a 
$|V_{ub}|$ measurement via relation (\ref{oldrelation}). Lastly, as
we will see below, the weight function $W(\Delta,P_+)$ possesses a
much better perturbative expansion than the function 
$\widetilde W(\Delta,P_+)=|V_{tb} V_{ts}^*|^2\,W(\Delta,P_+)/\Gamma_s(E_*)$. 
This last point can be traced back to the fact that most of the very large
contribution from the $O_1 - O_{7\gamma}$ operator mixing in the effective 
weak Hamiltonian cancels in the theoretical expression for the
normalized photon spectrum.

In principle, any partial $\bar B\to X_u\,l^-\bar\nu$ decay rate can
be brought into the form (\ref{eq:relation}), with complicated weight
functions. The relation between the two $P_+$ spectra is particularly simple, 
because the leading-power weight function is a constant at tree level. 
Experiments typically reject semileptonic $B$-decay events with very low 
lepton energy. The effect of such an additional cut can be determined from
\cite{Lange:2005yw}. Alternatively, it is possible to modify the weight
function so as to account for a lepton cut, however at the expense 
of a significant increase in complexity. We will not pursue this option in 
the present work.

The weight function depends on the kinematical variable $P_+$ and on the size 
$\Delta$ of the integration domain. It possesses integrable singularities of
the form $\alpha_s^n(\mu) \ln^k[m_b(\Delta-P_+)/\mu^2]$, with $k\le n$, in
perturbation theory. Different strategies can be found in the
literature concerning these logarithms. Leibovich et al.\ resummed them
by identifying $\mu^2$ with $m_b(\Delta-P_+)$
\cite{Leibovich:1999xf,Leibovich:2000ey}. The $P_+$ dependence of the
weight function then enters via the running coupling
$\alpha_s(\sqrt{m_b(\Delta-P_+)})$. This is a legitimate choice 
of scale as long as $(\Delta-P_+)$ has a generic value of order 
$\Lambda_{\rm QCD}$; however, it is {\em not\/} a valid choice in the small 
region where $m_b(\Delta-P_+)\sim\Lambda_{\rm QCD}^2$. A key result 
underlying relation (\ref{eq:relation}) is that, by construction, the weight 
function is insensitive to soft physics. Quark-hadron duality ensures that 
the region near the point $P_+=\Delta$, which is without any physical 
significance, does not require special consideration after 
integration over $P_+$. In the approach of 
\cite{Leibovich:1999xf,Leibovich:2000ey}, the attempt to resum the above 
logarithms near the endpoint of the $P_+$ integral
leads to integrals over unphysical Landau singularities of the running 
coupling in the nonperturbative domain. Hoang et al.\ chose to calculate the
weight function in fixed-order perturbation theory at the scale $\mu=m_b$ 
\cite{Hoang:2005pj}. This leads to parametrically large logarithms, 
since $\Delta-P_+\ll m_b$. In the present paper we separate physics effects 
from two parametrically distinct scales, a hard scale $\mu_h\sim m_b$ and
an intermediate scale $\mu_i\sim \sqrt{m_b\Lambda_{\rm QCD}}$, so that
we neither encounter Landau singularities nor introduce parametrically
large logarithms. The shape of the weight function is then governed by
a perturbative expansion at the intermediate scale, $\alpha_s^n(\mu_i)
\ln^k[m_b(\Delta-P_+)/\mu_i^2]$. As will be explained later, the coefficients 
in this series, as well as the overall normalization, possess themselves an
expansion in $\alpha_s(\mu_h)$.

The calculation of the weight function starts with the
theoretical expressions for the $P_+$ spectra in $\bar B\to X_u\,l^-\bar\nu$ 
and $\bar B\to X_s\gamma$ decays, which are given as \cite{Lange:2005yw}
\begin{eqnarray}
   \frac{d\Gamma_u}{dP_+} 
   &=& \frac{G_F^2 |V_{ub}|^2}{96\pi^3}\,U(\mu_h,\mu_i)\,(M_B-P_+)^5
    \label{eq:gammaU} \\
   &&\times \int_0^1\!dy\,y^{2-2a_\Gamma(\mu_h,\mu_i)} 
    \Big\{ (3-2y)\,\F_1(P_+,y) + 6(1-y)\,\F_2(P_+,y) + y\,\F_3(P_+,y) \Big\}
    \,, \nonumber\\
   \frac{d\Gamma_s}{dP_+} 
   &=& \frac{\alpha G_F^2 |V_{tb}V^*_{ts}|^2}{32\pi^4}\,U(\mu_h,\mu_i)\,
    (M_B-P_+)^3\,\overline{m}_b^2(\mu_h)\,[C_{7\gamma}^{\rm eff}(\mu_h)]^2\,
    \F_\gamma(P_+) \,, \label{eq:gammaS}
\end{eqnarray}
where 
\begin{equation} \label{eq:y}
   y = \frac{P_- -P_+}{M_B-P_+}
\end{equation}
with $P_-=E_X+|\vec{P}_X|$ 
is a kinematical variable that is integrated over the available phase space. 
Expressions for the structure functions $\F_i$ valid at next-to-leading order
(NLO) in renormalization-group (RG) improved perturbation theory and including 
first- and second-order power corrections can be found in \cite{Lange:2005yw} 
(see also \cite{Neubert:2004dd,Bauer:2003pi,Bosch:2004th}). Symbolically, they 
are written as $H(\mu_h)\cdot J(\mu_i)\otimes S(\mu_i)$, where $H(\mu_h)$ 
contains matching
corrections at the hard scale $\mu_h$. The jet function $J(\mu_i)$,
which is a perturbative quantity at the intermediate scale $\mu_i$, is
convoluted with a non-perturbative shape function renormalized at that same
scale. Separation of the two scales $\mu_h$ and $\mu_i$ allows for the
logarithms in matching corrections to be small, while logarithms of
the form $\ln\mu_h/\mu_i$, which appear at every
order in perturbation theory, are resummed in a systematic fashion and
give rise to the RG evolution functions $U(\mu_h,\mu_i)$ and
$a_\Gamma(\mu_h,\mu_i)$ \cite{Bosch:2004th}.

The leading-power jet function $J(p^2,\mu_i)$ entering the expressions for 
$\F_i$ is universal and has been computed at one-loop order in
\cite{Bauer:2003pi,Bosch:2004th}. More recently, the two-loop expression for 
$J$ has been obtained apart from a single unknown 
constant \cite{Neubert:2005nt}, which is the two-loop coefficient of 
the local $\delta(p^2)$ term.
This constant does not enter in the two-loop result for the weight 
function $W(\Delta,P_+)$ in (\ref{eq:relation}). Due to the
universality of the leading-power jet function, it is possible to
calculate the complete $O(\alpha_s^2(\mu_i))$ corrections to the weight
function. However, the extraction of hard corrections at two-loop order 
would require multi-loop calculations for both decay
processes, which are unavailable at present. As a result, we will be able to 
predict the $\Delta$ and $P_+$ dependence of the weight function 
$W(\Delta,P_+)$ at next-to-next-to-leading order (NNLO) in RG-improved 
perturbation theory, including exact two-loop matching contributions and 
three-loop 
running effects. However, the overall normalization of the weight function 
will have an uncertainty of $O(\alpha_s^2(\mu_h))$ from yet unknown hard 
matching corrections.

The total rate $\Gamma_s(E_*)$ has been calculated in a local operator
product expansion and reads (including only the leading non-perturbative 
corrections) \cite{Neubert:2004dd}
\begin{equation}\label{eq:totalBtoS}
   \Gamma_s(E_*) = \frac{\alpha G_F^2 |V_{tb}V^*_{ts}|^2}{32\pi^4}\,
   m_b^3\,\overline{m}_b^2(\mu_h)\,[C_{7\gamma}^{\rm eff}(\mu_h)]^2\,
   |H_s(\mu_h)|^2\,H_\Gamma(\mu_h) \left[ 1 - \frac{\lambda_2}{9m_c^2}\, 
   \frac{C_1(\mu_h)}{C_{7\gamma}^{\rm eff}(\mu_h)} \right] ,
\end{equation}
where $H_s$ is the hard function of $\F_\gamma$, and $H_\Gamma$ contains the 
remaining radiative corrections. We will present an explicit expression for 
this quantity at the end of Section~\ref{sec:results} below. The hadronic 
correction proportional to $\lambda_2/m_c^2$ cancels against an identical 
term in $\F_\gamma$. Apart from two 
powers of the running $b$-quark mass defined in the $\overline{\rm MS}$ 
scheme, which is part of the electromagnetic dipole operator $O_{7\gamma}$ in 
the effective weak Hamiltonian, three more powers of $m_b$ emerge from 
phase-space integrations. To avoid the renormalon
ambiguities of the pole scheme we use a low-scale subtracted quark-mass 
definition for $m_b$. Specifically, we adopt the shape-function mass 
$m_b^{\rm SF}(\mu_*,\mu_*)$ \cite{Bosch:2004th,Neubert:2004sp} defined at a 
subtraction scale $\mu_*=1.5$\,GeV, which relates to the pole mass as
\begin{equation}\label{eq:SFmassToPole}
   m_b^{\rm pole} = m_b^{\rm SF}(\mu_*,\mu_*)
   + \mu_*\,\frac{C_F \alpha_s(\mu_h)}{\pi} + \ldots \,.
\end{equation}
Throughout this paper we will use $m_b^{\rm SF}(\mu_*,\mu_*)$ as the
$b$-quark mass and refer to it as $m_b$ for brevity. The present value of 
this parameter is $m_b=(4.61\pm 0.06)$\,GeV \cite{Neubert:2005nt}.

\section{Calculation of the weight function}

\subsection{Leading power}

The key strategy for the calculation of the weight function is to make use of 
QCD factorization theorems for the decay distributions on both sides of
(\ref{eq:relation}) and to arrange the resulting, factorized expressions 
such that they are both given as integrals over the shape function, 
$\int d\hat\omega\,\hat S(\hat\omega)\,g_i(\hat\omega)$, with different 
functions $g_i$ for the
left-hand and right-hand sides. Relation~(\ref{eq:relation}) 
can then be enforced by matching $g_{\rm LHS}$ to $g_{\rm RHS}$. 
Following this procedure, we find for the integrated $P_+$ spectrum in
$\bar B\to X_u\,l^-\bar\nu$ decay after a series of integration interchanges
\begin{eqnarray}\label{eq:uSwitch}
   \int\limits_0^\Delta dP_+\,\frac{d\Gamma_u}{dP_+} 
   &\propto& \int\limits_0^1\!dy\,y^{-2a} H_u(y) 
    \int\limits_0^\Delta\!dP_+\,(M_B-P_+)^5
    \int\limits_0^{P_+}\!d\hat\omega\,y m_b\,J(ym_b(P_+ -\hat\omega))\,
    \hat S(\hat\omega) \nonumber \\
   &=& \int\limits_0^\Delta\!d\hat\omega\,\hat S(\hat\omega) 
    \int\limits_0^{M_B-\hat\omega}\!dq\,5q^4 
    \int\limits_0^1\!dy\,y^{-2a} H_u(y)\,
    j\left(\ln \frac{m_b(\Delta_q-\hat\omega)}{\mu_i^2} + \ln y \right) ,
\end{eqnarray}
where $\Delta_q=\mbox{min}(\Delta,M_B-q)$, and \cite{Neubert:2005nt}
\begin{equation}\label{jdef}
   j\left(\ln \frac{Q^2}{\mu_i^2},\mu_i \right)
   = \int_0^{Q^2}\!dp^2\,J(p^2,\mu_i)
\end{equation}
is the integral over the jet function. For the sake of transparency, we 
often suppress the explicit dependence on $\mu_i$ and $\mu_h$ when it is clear 
at which scales the relevant quantities are defined. The function $H_u$
is a linear combination of the hard functions entering the structures
$\F_i$ in (\ref{eq:gammaU}), which in the notation of \cite{Lange:2005yw} is
given by
\begin{equation}
   H_u(y,\mu_h) = 2y^2(3-2y)\,H_{u1}(y,\mu_h) + 12y^2(1-y)\,H_{u2}(y,\mu_h) 
   + 2y^3\,H_{u3}(y,\mu_h) \,.
\end{equation}
RG resummation effects build up the factor $y^{-2a}$ in (\ref{eq:uSwitch}), 
where $a\equiv a_\Gamma(\mu_h,\mu_i)$ is the value of the RG-evolution 
function
\begin{equation}\label{adef}
   a_\Gamma(\mu_h,\mu_i)
   = \int_{\mu_i}^{\mu_h}\!\frac{d\mu}{\mu}\,\Gamma_{\rm cusp}(\alpha_s(\mu))
   = - \int_{\alpha_s(\mu_h)}^{\alpha_s(\mu_i)}\!d\alpha\,
    \frac{\Gamma_{\rm cusp}(\alpha)}{\beta(\alpha)} \,,
\end{equation}
which depends only on the cusp anomalous dimension 
\cite{Korchemsky:wg,Korchemskaya:1992je}. The quantity $a$ has its origin in 
the geometry of time-like and light-like Wilson lines underlying the 
kinematics of inclusive $B$ decays into light particles. Our definition is 
such that $a$ is a positive number for $\mu_h>\mu_i$ and vanishes in the limit 
$\mu_h\to\mu_i$. We find it convenient to treat the function 
$a_\Gamma(\mu_h,\mu_i)$ as a running ``physical'' 
quantity, much like $\alpha_s(\mu)$ or $\overline{m}_b(\mu)$. Since the 
cusp anomalous dimension is known to three-loop order \cite{Moch:2004pa}, the 
value of $a$ can be determined very accurately. Note that three-loop accuracy 
in $a$ (as well as in the running coupling $\alpha_s$) is required for a 
consistent calculation of the weight function at NNLO. The corresponding 
expression is
\begin{eqnarray}
   a = a_\Gamma(\mu_h,\mu_i) &=& \frac{\Gamma_0}{2\beta_0}\,\Bigg\{
    \ln\frac{\alpha_s(\mu_i)}{\alpha_s(\mu_h)} 
    + \left( \frac{\Gamma_1}{\Gamma_0} - \frac{\beta_1}{\beta_0} \right)
    \frac{\alpha_s(\mu_i) - \alpha_s(\mu_h)}{4\pi} \nonumber\\
   &&\mbox{}+ \left[ \frac{\Gamma_2}{\Gamma_0} - \frac{\beta_2}{\beta_0}
    - \frac{\beta_1}{\beta_0}
    \left( \frac{\Gamma_1}{\Gamma_0} - \frac{\beta_1}{\beta_0} \right) \right]
    \frac{\alpha_s^2(\mu_i) - \alpha_s^2(\mu_h)}{32\pi^2} + \dots \Bigg\} \,,
\end{eqnarray}
where the expansion coefficients $\Gamma_n$ and $\beta_n$ of the cusp anomalous
dimension and $\beta$-function can be found, e.g., in \cite{Neubert:2004dd}.

Instead of the jet function $J$ itself, we need its integral 
$j(\ln Q^2/\mu_i^2,\mu_i)$ in the second line of (\ref{eq:uSwitch}).
Since the jet function has a
perturbative expansion in terms of ``star distributions'', which are
logarithmically sensitive to the upper limit of integration
\cite{DeFazio:1999sv}, it follows that $j(L,\mu_i)$ is a simple polynomial in
$L$ at each order in perturbation theory. The two-loop result for this
quantity has recently been computed by solving the integro-differential 
evolution equation for the jet function \cite{Neubert:2005nt}. An unknown 
integration constant of $O(\alpha_s^2 L^0)$ does not enter the expression for 
the weight function.

We now turn to the right-hand side of (\ref{eq:relation}) and follow
the same steps that lead to (\ref{eq:uSwitch}). It is helpful to make
an ansatz for the leading-power contribution to the weight function, 
$W^{(0)}(\Delta,P_+)$, where the 
dependence on $\Delta$ is solely given via an upper limit of integration. To 
this end, we define a function $f(k)$ through
\begin{equation}\label{eq:fdef}
   W^{(0)}(\Delta,P_+)\propto \frac{1}{(M_B-P_+)^3} 
   \int_0^{\Delta-P_+}\!dk\,f(k)\,(M_B-P_+-k)^5 \,.
\end{equation}
This allows us to express the weighted integral over the $\bar B\to X_s\gamma$ 
photon spectrum as
\begin{eqnarray}\label{eq:sSwitch}
   \int\limits_0^\Delta\!dP_+\,\frac{d\Gamma_s}{dP_+}\,W^{(0)}(\Delta,P_+)
   &\propto& \int\limits_0^\Delta\!dP_+\,(M_B-P_+)^3\,W^{(0)}(\Delta,P_+)
    \int\limits_0^{P_+}\!d\hat\omega\,m_b\,J(m_b(P_+ -\hat\omega))\,
    \hat S(\hat\omega) \nonumber \\ 
   &\propto& \int\limits_0^\Delta\!d\hat\omega\,\hat S(\hat\omega) 
    \int\limits_0^{M_B-\hat\omega}\!dq\,5q^4 
    \int\limits_0^{\Delta_q-\hat\omega}\!dk\,f(k)\, 
    j\left(\ln\frac{m_b(\Delta_q-\hat\omega-k)}{\mu_i^2} \right) . \qquad
\end{eqnarray}
Note that the jet function $J$ (and with it $j$) is the same in semileptonic 
and radiative decays. The difference is that the argument of the jet 
function in (\ref{eq:uSwitch}) contains an extra factor of $y$, which is 
absent in (\ref{eq:sSwitch}). Comparing these two relations leads us to the 
matching condition
\begin{equation} \label{eq:fmatch}
   \int_0^1\!dy\, y^{-2a}\,H_u(y)\,
   j\left(\ln\frac{m_b\Omega}{\mu_i^2}+\ln y \right)
   \stackrel{!}{=} \int_0^\Omega\!dk\,f(k)\,
   j\left(\ln\frac{m_b(\Omega-k)}{\mu_i^2} \right) ,
\end{equation}
which holds to all orders in perturbation theory and allows for the
calculation of $W^{(0)}(\Delta,P_+)$ via (\ref{eq:fdef}). The main feature
of this important relation is that the particular value of
$\Omega$ is irrelevant for the determination of $f(k)$. It follows
that, as was the case for the jet function $J$, the perturbative
expansion of $f(k)$ in $\alpha_s(\mu_i)$ at the intermediate scale
involves star distributions, and $W^{(0)}(\Delta,P_+)$ depends
logarithmically on $(\Delta-P_+)$. At two-loop order it suffices to
make the ansatz
\begin{eqnarray}\label{eq:fexpansion}
   f(k) &\propto& \delta(k) + C_F\frac{\alpha_s(\mu_i)}{4\pi}
    \left[ c_0^{(1)}\,\delta(k) 
    + c_1^{(1)} \left( \frac{1}{k} \right)_*^{[\mu_i^2/m_b]} \right] \\
   &&\mbox{}+ C_F\left( \frac{\alpha_s(\mu_i)}{4\pi} \right)^2 
    \left[ c_0^{(2)}\,\delta(k)
    + c_1^{(2)} \left( \frac{1}{k} \right)_*^{[\mu_i^2/m_b]}
    + 2c_2^{(2)} \left( \frac{1}{k}\ln\frac{m_b k}{\mu_i^2} 
    \right)_*^{[\mu_i^2/m_b]} \right] + \dots \,, \nonumber 
\end{eqnarray}
where the star distributions have the following effect when integrated
with some smooth function $\phi(k)$ over an interval $\Omega$:
\begin{eqnarray}\label{stardistris}
   \int_0^\Omega\!dk \left( \frac{1}{k} \right)_*^{[\mu_i^2/m_b]} \phi(k)
   &=& \int_0^{\Omega}\!dk\,\frac{\phi(k)-\phi(0)}{k}
    + \phi(0)\,\ln\frac{m_b\Omega}{\mu_i^2} \,, \nonumber\\
   \int_0^\Omega\!dk\,\left( \frac{1}{k}\ln\frac{m_b k}{\mu_i^2} 
    \right)_*^{[\mu_i^2/m_b]} \phi(k)
   &=& \int_0^{\Omega}\!dk\,\frac{\phi(k)-\phi(0)}{k}\,
    \ln\frac{m_b k}{\mu_i^2} \,
    + \frac{\phi(0)}{2}\,\ln^2\frac{m_b\Omega}{\mu_i^2} \,.
\end{eqnarray}

A sensitivity to the hard scale $\mu_h$ enters into $f(k)$ via the
appearance of $H_u(y,\mu_h)$ in (\ref{eq:fmatch}). Because of the
polynomial nature of $j(L,\mu_i)$, all we ever need are moments of the hard 
function with respect to $\ln y$. We thus define the master integrals
\begin{equation}\label{eq:Tn}
   T_n(a,\mu_h)\equiv \int_0^1\!dy\,y^{-2a}\,H_u(y,\mu_h)\,\ln^n y \,; \qquad 
   h_n(a,\mu_h) = \frac{T_n(a,\mu_h)}{T_0(a,\mu_h)} \,,
\end{equation}
which can be calculated order by order in $\alpha_s(\mu_h)$. Therefore,
the coefficients $c_k^{(n)}$ of the perturbative expansion in
(\ref{eq:fexpansion}) at the intermediate scale have the (somewhat
unusual) feature that they possess themselves an expansion in
$\alpha_s(\mu_h)$. This is a consequence of the fact that, unlike the
differential decay rates (\ref{eq:gammaU}) and (\ref{eq:gammaS}), the
weight function itself does not obey a simple factorization
formula, in which the hard correction can be factored out. Rather, as can be 
seen from (\ref{eq:fmatch}), it is a convolution of the type 
$W=H(\mu_h)\otimes J(\mu_i)$. 
To one-loop accuracy, the hard function $H_u$ reads
\begin{eqnarray}\label{eq:Hu}
   H_u(y,\mu_h) 
   &=& 2y^2(3-2y)\,\Bigg[ 1 + \frac{C_F\alpha_s(\mu_h)}{4\pi}\,
    \bigg( -4\ln^2\frac{ym_b}{\mu_h} + 10\ln\frac{ym_b}{\mu_h} - 4\ln y 
    \nonumber\\
   &&\hspace{2.3cm}\mbox{}- 4 L_2(1-y) - \frac{\pi^2}{6} - 12 \bigg) \Bigg] 
    - \frac{C_F\alpha_s(\mu_h)}{\pi}\,3y^2\ln y \,.
\end{eqnarray}
Explicit expressions for the quantities $T_0$, $c_k^{(n)}$, and $h_n$
entering the distribution function $f(k)$ will be given below.

\subsection{Subleading power}
\label{sec:corr}

Power corrections to the weight function can be 
extracted from the corresponding contributions to the two $P_+$ spectra in 
(\ref{eq:gammaU}) and (\ref{eq:gammaS}). There exists a class of power 
corrections associated with the phase-space prefactors $(M_B-P_+)^n$ in these 
relations, whose effects are treated exactly in our approach, see e.g.\ 
(\ref{eq:fdef}). This is 
important, because these phase-space corrections increase in magnitude as the 
kinematical range $\Delta$ over which the two spectra are integrated is 
enlarged. One wants to make $\Delta$ as large as 
experimentally possible so as to increase statistics and justify the 
assumption of quark-hadron duality, which underlies the theory of inclusive
$B$ decays. 

The remaining power corrections fall into two distinct classes:
kinematical corrections that start at order $\alpha_s$ and come with  
the leading shape function \cite{Kagan:1998ym,DeFazio:1999sv}, and 
hadronic power corrections that start at tree level and involve new, 
subleading shape functions 
\cite{Bauer:2001mh,Leibovich:2002ys,Bauer:2002yu,Neubert:2002yx,%
Burrell:2003cf,Lee:2004ja,Bosch:2004cb,Beneke:2004in}. Because different 
combinations of these hadronic functions enter in $\bar B\to X_u\,l^-\bar\nu$ 
and $\bar B\to X_s\gamma$ decays, it is impossible to eliminate their 
contributions in relations such as (\ref{eq:relation}). As a result, at 
$O(\Lambda_{\rm QCD}/m_b)$ there are non-perturbative hadronic uncertainties
in the calculation of the weight function $W(\Delta,P_+)$, which need to be
estimated before a reliable extraction of $|V_{ub}|$ can be performed. For
the case of the charged-lepton energy spectrum and the hadronic invariant mass
spectrum, this aspect has been discussed previously in 
\cite{Bauer:2002yu,Neubert:2002yx} and \cite{Burrell:2003cf}, respectively.

Below, we will include power corrections to first order in 
$\Lambda_{\rm QCD}/m_b$. Schematically, the subleading corrections to the 
right-hand side of (\ref{eq:relation}) are computed according to
$\Gamma_u^{(1)}\sim W^{(0)}\otimes d\Gamma_s^{(1)}/dP_+%
+W^{(1)}\otimes d\Gamma_s^{(0)}/dP_+$, where the superscripts indicate the
order in $1/m_b$ power counting. The power corrections to the weight function, 
denoted by $W^{(1)}$, are derived from the mismatch in the power corrections 
to the two decay spectra. The kinematical power corrections to the two spectra 
are known at $O(\alpha_s)$, without scale separation. We assign a coupling 
$\alpha_s(\bar\mu)$ to these terms, where the scale $\bar\mu$ will be chosen
of order the intermediate scale \cite{Lange:2005yw}. At first subleading power 
the leading shape function is convoluted with either a constant or a
single logarithm of the form $\ln[(P_+ -\hat\omega)/(M_B-P_+)]$, and we have
(with $n=0,1$)
\begin{eqnarray}
   \int_0^\Delta\!dP_+\,\frac{d\Gamma_u}{dP_+} 
   &\ni& \alpha_s(\bar\mu) \int_0^\Delta\!dP_+\,(M_B-P_+)^4
    \int_0^{P_+}\!d\hat\omega\,\hat S(\hat\omega)\,
    \ln^n\frac{P_+ -\hat\omega}{M_B-P_+} \nonumber \\
   &=& \alpha_s(\bar\mu) \int_0^\Delta\!d\hat\omega\,\hat S(\hat\omega) 
    \int_0^{\Delta-\hat\omega}\!dk\,(M_B-\hat\omega-k)^4 
    \ln^n\frac{k}{M_B-\hat\omega-k} \,,
\end{eqnarray}
and similarly for the photon spectrum. On the other hand, the weighted
integral in (\ref{eq:relation}) also contains terms where the photon
spectrum is of leading power and the weight function of subleading
power,
\begin{equation}
   \int_0^\Delta\!dP_+\,\frac{d\Gamma_s}{dP_+}\,W^{\rm kin(1)}(\Delta,P_+)
   \ni \int_0^\Delta\!dP_+\,(M_B-P_+)^3\,\hat S(P_+)\,
   W^{\rm kin(1)}(\Delta,P_+) \,.
\end{equation}
Therefore the kinematical corrections to the weight function must have the 
form 
\begin{equation}\label{eq:wkin1}
   W^{\rm kin(1)}(\Delta,P_+)\propto \frac{\alpha_s(\bar\mu)}{(M_B-P_+)^3} 
   \int_0^{\Delta-P_+}\!dk\,(M_B-P_+ -k)^4 
   \left( A + B\,\ln\frac{k}{M_B-P_+ -k} \right) ,
\end{equation}
and a straightforward calculation determines the coefficients $A$ and $B$.  
The hadronic power corrections to the weight function, $W^{\rm hadr(1)}$, 
can be expressed in terms of the subleading shape functions 
$\hat t(\hat\omega)$, $\hat u(\hat\omega)$, and $\hat v(\hat\omega)$ defined 
in \cite{Bosch:2004cb}. These terms are known at tree level only, and at
this order their contribution to the weight function can be derived using the 
results of \cite{Lange:2005yw}.

\section{Results}
\label{sec:results}

Including the first-order power corrections and the exact phase-space
factors, the weight function takes the form
\begin{eqnarray}\label{eq:masterform}
   W(\Delta,P_+)
   &=& \frac{G_F^2 m_b^3}{192\pi^3}\,T_0(a,\mu_h)\,H_\Gamma(\mu_h)\,
    (M_B-P_+)^2 \nonumber\\
   &\times& \Bigg\{ 1 + \frac{C_F\alpha_s(\mu_i)}{4\pi}
    \left[ c_0^{(1)} + c_1^{(1)} \left( \ln \frac{m_b(\Delta-P_+)}{\mu_i^2} 
    - p_1(\delta) \right) \right] \nonumber\\
   &&\hspace{0.55cm}\mbox{}+ C_F \left( \frac{\alpha_s(\mu_i)}{4\pi} \right)^2
    \left[ c_0^{(2)} + c_1^{(2)} \left( \ln \frac{m_b(\Delta-P_+)}{\mu_i^2} 
    - p_1(\delta) \right) \right. \nonumber\\
   &&\hspace{1.6cm}\left.
    \mbox{}+ c_2^{(2)} \left( \ln^2\frac{m_b(\Delta-P_+)}{\mu_i^2} 
    - 2 p_1(\delta) \ln\frac{m_b(\Delta-P_+)}{\mu_i^2} 
    + 2 p_2(\delta) \right) \right] \nonumber\\
   &&\mbox{}+ \frac{\Delta-P_+}{M_B-P_+}\,\frac{C_F\alpha_s(\bar\mu)}{4\pi}
    \left[ A(a,\mu_h)\,I_A(\delta) + B(a,\mu_h)\,I_B(\delta) \right] \\
   &&\mbox{}+ \frac{1}{M_B-P_+}\,\frac{1}{2(1-a)(3-a)}\,
    \bigg[ 4(1-a)(\bar\Lambda-P_+) + 2(4-3a)\,\frac{\hat t(P_+)}{\hat S(P_+)}
    \nonumber\\
   &&\mbox{}+ (4-a)\,\frac{\hat u(P_+)}{\hat S(P_+)}
    + (8-13a+4a^2)\,\frac{\hat v(P_+)}{\hat S(P_+)} \bigg]
    + \frac{m_s^2}{M_B-P_+}\,\frac{\hat S'(P_+)}{\hat S(P_+)} 
    + \dots \Bigg\} \,, \nonumber
\end{eqnarray}
where $\bar\Lambda=M_B-m_b$ is the familiar mass parameter of heavy-quark 
effective theory, and $\delta=(\Delta-P_+)/(M_B-P_+)$. 
The first line denotes an overall normalization, 
the next three lines contain the leading-power contributions, and
the remaining expressions enter at subleading power. The different terms in
this result will be discussed in the remainder of this section.

For the leading-power terms in the above result we have accomplished a 
complete separation of hard and intermediate (hard-collinear) contributions 
to the weight function in a way consistent with the factorization formula
$W=H\otimes J$ mentioned in the previous section. The 
universality of the shape function, which encodes the soft physics in both 
$\bar B\to X_s\gamma$ and $\bar B\to X_u\,l^-\bar\nu$ decays, implies that 
the weight function is insensitive to physics below the intermediate scale
$\mu_i\sim\sqrt{m_b\Lambda_{\rm QCD}}$. In particular, quark-hadron duality 
ensures that the small region in phase space where the argument 
$m_b(\Delta-P_+)$ of the logarithms scales as $\Lambda_{\rm QCD}^2$ or 
smaller does not need special consideration. 
At a technical level, this can be seen by noting that the jet function is the
discontinuity of the collinear quark propagator in soft-collinear effective 
theory \cite{Bosch:2004th,Bauer:2001yt}, and so the $P_+$ integrals can be 
rewritten as a contour integral in the complex $p^2$ plane along a circle of 
radius $m_b\Delta\sim\mu_i^2$.

\subsection{Leading power}

The leading-power corrections in the curly brackets in (\ref{eq:masterform})
are determined completely 
at NNLO in RG-improved perturbation theory, including three-loop running 
effects via the quantity $a$ in (\ref{adef}), and two-loop matching 
corrections at the scale $\mu_i$ as indicated above. To this end we need
expressions for the one-loop coefficients $c_n^{(1)}$ including terms of
$O(\alpha_s(\mu_h))$, while the two-loop coefficients $c_n^{(2)}$ are needed 
at leading order only. We find
\begin{equation}\label{eq:1-loopCoeffs}
   c_0^{(1)} = -3 h_1(a,\mu_h) 
    \left[ 1 - \frac{C_F\alpha_s(\mu_h)}{\pi}\,\frac{4\mu_*}{3m_b} \right]
    + 2 h_2(a,\mu_h) \,, \qquad
   c_1^{(1)} = 4 h_1(a,\mu_h) \,,
\end{equation}
and 
\begin{eqnarray}\label{eq:2-loopCoeffs}
   c_0^{(2)}
   &=& C_F\,\Bigg[ \left( -\frac32 + 2\pi^2 - 24\zeta_3 \right) h_1(a,\mu_h) 
    + \left( \frac92 - \frac{4\pi^2}{3} \right) h_2(a,\mu_h) 
    - 6 h_3(a,\mu_h) + 2 h_4(a,\mu_h) \Bigg] \nonumber\\
   &&\mbox{}+ C_A \left[ \left( -\frac{73}{9} + 40\zeta_3 \right) h_1(a,\mu_h)
    + \left( \frac83 - \frac{2\pi^2}{3} \right) h_2(a,\mu_h) \right] 
    \nonumber\\
   &&\mbox{}+ \beta_0 \left[
    \left( -\frac{247}{18} + \frac{2\pi^2}{3} \right) h_1(a,\mu_h)
    + \frac{29}{6}\,h_2(a,\mu_h) - \frac23\,h_3(a,\mu_h) \right] , \nonumber\\
   c_1^{(2)} 
   &=& C_F\,\Big[ -12 h_2(a,\mu_h) + 8 h_3(a,\mu_h) \Big]
    + C_A \left[ \left( \frac{16}{3} - \frac{4\pi^2}{3} \right) h_1(a,\mu_h)
    \right] \nonumber\\
   &&\mbox{}+ \beta_0 \left[ \frac{29}{3}\,h_1(a,\mu_h) - 2 h_2(a,\mu_h)
    \right] , \nonumber \\
   c_2^{(2)} &=& 8 C_F\,h_2(a,\mu_h) - 2\beta_0\,h_1(a,\mu_h) \,.
\end{eqnarray}
As always $C_F=4/3$, $C_A=3$, and $\beta_0=11-2 n_f/3$ is the first 
coefficient of the QCD $\beta$-function. The term proportional to $\mu_*$ in 
the expression for $c_0^{(1)}$ arises because of the elimination of the 
pole mass in favor of the shape-function mass, see
(\ref{eq:SFmassToPole}). Since the logarithms $\ln[m_b(\Delta-P_+)/\mu_i^2]$ 
in (\ref{eq:masterform}) contain $m_b$, all coefficients except $c_n^{(n)}$ 
receive such contributions. However, to two-loop order only $c_0^{(1)}$ is 
affected.

Next, the corresponding expressions for the hard matching coefficients
$h_i$ are calculated from (\ref{eq:Tn}). To the required order they read
\begin{eqnarray}
   h_1(a,\mu_h)
   &=& - \frac{15-12a+2a^2}{2(2-a)(3-a)(3-2a)} \nonumber\\
   &&\mbox{}+ \frac{C_F\alpha_s(\mu_h)}{4\pi}\,\Bigg[
    - \frac{2(189-318a+192a^2-48a^3+4a^4)}{(2-a)^2(3-a)^2(3-2a)^2}\,
    \ln\frac{m_b}{\mu_h} \nonumber\\
   &&\quad\mbox{}+
    \frac{2331-5844a+5849a^2-2919a^3+726a^4-72a^5}{(2-a)^3(3-a)^2(3-2a)^3}
    - 4\psi^{(2)}(3-2a) \Bigg] + \dots \,, \nonumber
\end{eqnarray}
\begin{eqnarray}
   h_2(a,\mu_h)
   &=&  \frac{69-90a+36a^2-4a^3}{2(2-a)^2(3-a)(3-2a)^2} \nonumber\\
   &&\mbox{}+ \frac{C_F\alpha_s(\mu_h)}{4\pi}\,\Bigg[
    \frac{2(1692-3699a+3138a^2-1272a^3+240a^4-16a^5)}{(2-a)^3(3-a)^2(3-2a)^3}
    \,\ln\frac{m_b}{\mu_h} \nonumber\\
   &&\quad\mbox{}-
    \frac{46521-140064a+175479a^2-117026a^3+43788a^4-8712a^5+720a^6}
         {2(2-a)^4(3-a)^2(3-2a)^4} \nonumber\\
   &&\quad\mbox{}+ \frac{4(15-12a+2a^2)}{(2-a)(3-a)(3-2a)}\,\psi^{(2)}(3-2a)
    - 4\psi^{(3)}(3-2a) \Bigg] + \dots \,, \nonumber\\
   h_3(a,\mu_h)
   &=& - \frac{3(303-552a+360a^2-96a^3+8a^4)}{4(2-a)^3(3-a)(3-2a)^3} + \dots
    \,, \nonumber\\
   h_4(a,\mu_h)
   &=& \frac{3(1293-3030a+2760a^2-1200a^3+240a^4-16a^5)}{2(2-a)^4(3-a)(3-2a)^4}
    + \dots \,,
    \phantom{aaaaaaaaaaaaaaaa}
\end{eqnarray}
where $\psi^{(n)}(x)$ is the $n$-th derivative of the polygamma
function. Because of the exact treatment of the phase space there are
corrections to the logarithms in (\ref{eq:masterform}), which are 
finite-order polynomials in the small ratio
$\delta=(\Delta-P_+)/(M_B-P_+)$. Explicitly,
\begin{eqnarray}
   p_1(\delta) &=& 5\delta - 5\delta^2 + \frac{10}{3}\,\delta^3
    - \frac54\,\delta^4 + \frac15\,\delta^5 \,, \nonumber\\
   p_2(\delta) &=& 5\delta - \frac52\,\delta^2 + \frac{10}{9}\,\delta^3
    - \frac{5}{16}\,\delta^4 + \frac{1}{25}\,\delta^5 \,.
\end{eqnarray}
This concludes the discussion of the leading-power expression for the
weight function. 

\subsection{Subleading power}
\label{sec:power}

The procedure for obtaining the kinematical power corrections to the weight 
function has been discussed in Section~\ref{sec:corr}. For the coefficients 
$A$ and $B$ in (\ref{eq:masterform}) we find
\begin{eqnarray}
   A(a,\mu_h)
   &=& \frac{-388+702a-429a^2+123a^3-34a^4+8a^5}{2(1-a)^2(2-a)(3-a)(3-2a)} 
    + \left(\frac13- \frac49\,\ln\frac{m_b}{m_s} \right) 
    \frac{[C_{8g}^{\rm eff}(\mu_h)]^2}{[C^{\rm eff}_{7\gamma}(\mu_h)]^2}
    \nonumber\\
   &&- \frac{10}{3}\,
    \frac{C_{8g}^{\rm eff}(\mu_h)}{C^{\rm eff}_{7\gamma}(\mu_h)}
    + \frac83 \left( 
    \frac{C_1(\mu_h)}{C^{\rm eff}_{7\gamma}(\mu_h)} 
    - \frac13\,\frac{C_1(\mu_h)\,C_{8g}^{\rm eff}(\mu_h)}
    {[C^{\rm eff}_{7\gamma}(\mu_h)]^2}\,\right) g_1(z)
    - \frac{16}{9}\,\frac{[C_1(\mu_h)]^2}{[C^{\rm eff}_{7\gamma}(\mu_h)]^2}\, 
    g_2(z) \,, \nonumber\\
   B(a,\mu_h) 
   &=& - \frac{2(8+a)}{(1-a)(3-a)}
    - \frac29\,\frac{[C_{8g}^{\rm eff}(\mu_h)]^2}
    {[C^{\rm eff}_{7\gamma}(\mu_h)]^2} \,.
\end{eqnarray}
Here $C_i(\mu_h)$ denote the (effective) Wilson coefficients of the
relevant operators in the effective weak Hamiltonian, which are real
functions in the Standard Model. The variable $z=(m_c/m_b)^2$ enters
via charm-loop penguin contributions to the hard function of the 
$\bar B\to X_s\gamma$ photon spectrum \cite{Kagan:1998ym}, and
\begin{equation}
   g_1(z) = \int_0^1\!dx\,x\,\mbox{Re} \left[\,
    \frac{z}{x}\,G\!\left(\frac{x}{z}\right) + \frac12 \,\right] , \qquad
   g_2(z) = \int_0^1\!dx\,(1-x) \left|\,\frac{z}{x}\,
    G\!\left(\frac{x}{z}\right) + \frac12\,\right|^2 ,
\end{equation}
with
\begin{equation}
   G(t) = \left\{ \begin{array}{ll}
   -2\arctan^2\!\sqrt{t/(4-t)} & ;~ t<4 \,, \\[0.1cm]
   2 \left( \ln\!\Big[(\sqrt{t}+\sqrt{t-4})/2\Big]
   - \displaystyle\frac{i\pi}{2} \right)^2 & ;~ t\ge 4 \,.
  \end{array} \right.
\end{equation}
Furthermore we need the integrals over $k$ in
(\ref{eq:wkin1}), which encode the phase-space corrections. They give rise to 
the functions
\begin{eqnarray}
   I_A(\delta) &=& 1 - 2\delta + 2\delta^2 - \delta^3 + \frac15\,\delta^4
    \,, \nonumber\\
   I_B(\delta) &=& I_A(\delta)\,\ln\frac{\delta}{1-\delta}
    + \frac{\ln(1-\delta)}{5\delta} - \frac45 + \frac35\,\delta
    - \frac{4}{15}\,\delta^2 + \frac{1}{20}\,\delta^3 \,.
\end{eqnarray}

The hadronic power corrections come from subleading shape functions
in the theoretical expressions for the two decay rates. We give their
tree-level contributions to the weight function in the last two lines
of (\ref{eq:masterform}), where $\hat S$ denotes the leading
shape function, and $\hat t, \hat u,\hat v$ are subleading shape
functions as defined in \cite{Bosch:2004cb}. For completeness, we also include 
a contribution proportional to $m_s^2$ resulting from finite-mass effects 
in the strange-quark propagator in $\bar B\to X_s\gamma$ decays. For 
$m_s=O(\Lambda_{\rm QCD})$ these effects are formally of the same order as 
other subleading shape-function contributions \cite{Chay:2005ck}, although
numerically they are strongly suppressed. The appearance of subleading shape
functions introduces an irreducible hadronic uncertainty to a $|V_{ub}|$
determination via (\ref{eq:relation}). In practice, this uncertainty
can be estimated by adopting different models for the subleading shape
functions. This will be discussed in detail in Section~\ref{sec:SSF} below. 
Until then, let us use a ``default model'', in which we 
assume the functional forms of the subleading shape functions 
$\hat t(\hat\omega)$, $\hat u(\hat\omega)$, and $\hat v(\hat\omega)$ to be 
particular linear combinations of the functions $\hat S'(\hat\omega)$ and 
$(\bar\Lambda-\hat\omega)\,\hat S(\hat\omega)$. These combinations are chosen 
in such a way that the 
results satisfy the moment relations derived in \cite{Bosch:2004cb}, and that 
all terms involving the parameter $\bar\Lambda$ cancel in the expression 
(\ref{eq:masterform}) for the weight function for any value of $a$. These 
requirements yield
\begin{equation}\label{defaultmodel}
   \hat t\to-\frac34\,(\bar\Lambda-\hat\omega)\,\hat S
    - \left( \lambda_2 + \frac{\lambda_1}{4} \right) \hat S' \,, \quad
   \hat u\to\frac12\,(\bar\Lambda-\hat\omega)\,\hat S
    + \frac{5\lambda_1}{6}\,\hat S' \,, \quad
   \hat v\to\lambda_2\,\hat S' \,,
\end{equation}
and the last two lines inside the large bracket in the expression 
(\ref{eq:masterform}) for the weight function simplify to
\begin{equation}\label{eq:SSFdefault}
   - \frac{\Lambda_{\rm SSF}^2(a)}{M_B-P_+}\,
   \frac{\hat S'(P_+)}{\hat S(P_+)} 
   ~\widehat{=} - \frac{\Lambda_{\rm SSF}^2(a)}{M_B-P_+}\,\delta(P_+ -\Delta)
    - \frac{4\Lambda_{\rm SSF}^2(a)}{(M_B-P_+)^2} \,,
\end{equation}
where 
\begin{equation}
   \Lambda_{\rm SSF}^2(a)\equiv - \frac{(2+a)\,\lambda_1}{3(1-a)(3-a)}
   + \frac{a(7-4a)\,\lambda_2}{2(1-a)(3-a)} - m_s^2 \,.
\end{equation}
Here $\lambda_1$ and $\lambda_2=\frac14(M_{B^*}^2-M_B^2)$ are hadronic 
parameters describing certain $B$-meson matrix elements in heavy-quark 
effective theory \cite{Falk:1992wt}. The strange-quark mass is a running 
mass evaluated at a scale typical for the final-state hadronic jet, for which 
we take 1.5\,GeV.
As mentioned above, the numerical effect of the strange-quark mass correction 
is small. For typical values of the parameters, it reduces the result for
$\Lambda_{\rm SSF}^2$ by about 10\% or less.
The expression on the right-hand side in (\ref{eq:SSFdefault}) is equivalent
to that on the left-hand side after the integration with the photon spectrum
in (\ref{eq:relation}) has been performed. It has been derived using the fact 
that the normalized photon spectrum is proportional to the shape function 
$\hat S(P_+)$ at leading order. Note that the second term in the final formula 
is power suppressed with respect to the first one. It results from our exact
treatment of phase-space factors and thus is kept for consistency.

\subsection{Normalization}

Finally, let us present explicit formulae for the overall normalization factor 
in (\ref{eq:masterform}). The new ingredient here is the factor $T_0$,
which is defined in (\ref{eq:Tn}). At one-loop order we find
\begin{eqnarray}
   T_0(a,\mu_h) 
   &=& \frac{2(3-a)}{(2-a)(3-2a)}\,\Bigg\{ 1
    - \frac{C_F\alpha_s(\mu_h)}{4\pi} \Bigg[ 4 \ln^2 \frac{m_b}{\mu_h}
    - \frac{2(120-159a+69a^2-10a^3)}{(2-a)(3-a)(3-2a)}\,\ln \frac{m_b}{\mu_h}
    \nonumber \\
   &&\mbox{}+
    \frac{1539-3845a+3842a^2-1920a^3+480a^4-48a^5}{(2-a)^2(3-a)(3-2a)^2} 
    + 4 \psi^{(1)}(3-2a) + \frac{\pi^2}{6} \Bigg] \Bigg\} \,. \nonumber\\
\end{eqnarray}
When the product of $T_0$ with the quantity \cite{Neubert:2004dd}
\begin{eqnarray}\label{Hgamma}
   H_\Gamma(\mu_h)
   &=& 1 + \frac{C_F\alpha_s(\mu_h)}{4\pi}\,
    \Bigg[ 4\ln^2\frac{m_b}{\mu_h} - 10\ln\frac{m_b}{\mu_h} 
    + 7 - \frac{7\pi^2}{6} + \frac{12\mu_*}{m_b} \nonumber \\
   &&\mbox{}- 2\ln^2\delta_* - (7+4\delta_*-\delta_*^2)\,\ln\delta_* 
    + 10\delta_* + \delta_*^2 - \frac23\,\delta_*^3 \nonumber \\
   &&\mbox{}+ \frac{[C_1(\mu_h)]^2}{[C^{\rm eff}_{7\gamma}(\mu_h)]^2}\, 
    \hat f_{11}(\delta_*)
    + \frac{C_1(\mu_h)}{C^{\rm eff}_{7\gamma}(\mu_h)}\,\hat f_{17}(\delta_*)
    + \frac{C_1(\mu_h)\,C_{8g}^{\rm eff}(\mu_h)}%
           {[C^{\rm eff}_{7\gamma}(\mu_h)]^2}\,\hat f_{18}(\delta_*)
    \nonumber\\
   &&\mbox{}+ \frac{C_{8g}^{\rm eff}(\mu_h)}{C^{\rm eff}_{7\gamma}(\mu_h)}\,
    \hat f_{78}(\delta_*)
    + \frac{[C_{8g}^{\rm eff}(\mu_h)]^2}{[C^{\rm eff}_{7\gamma}(\mu_h)]^2}\,
    \hat f_{88}(\delta_*) \Bigg]
\end{eqnarray}
from the total $\bar B\to X_s\gamma$ decay rate is consistently expanded to 
$O(\alpha_s(\mu_h))$, the double logarithm cancels out. Here 
$\delta_*=1-2E_*/m_b=0.9$, and the functions $\hat f_{ij}(\delta_*)$ capture 
effects from operator mixing.

\section{Numerical results}
\label{sec:num}

We are now in a position to explore the phenomenological implications of our
results. We need as inputs the heavy-quark 
parameters $\lambda_2=0.12$\,GeV$^2$, $\lambda_1=(-0.25\pm 0.10)$\,GeV$^2$, 
and the quark masses $m_b=(4.61\pm 0.06)$\,GeV \cite{Neubert:2005nt}, 
$m_s=(90\pm 25)$\,MeV \cite{Gamiz:2004ar,Aubin:2004ck}, and
$m_c/m_b=0.222\pm 0.027$ \cite{Neubert:2004dd}. Here $m_b$ is defined in the 
shape-function scheme at a scale $\mu_*=1.5$\,GeV, $m_s$ is the running mass 
in the $\overline{\rm MS}$ scheme evaluated at 1.5\,GeV, and $m_c/m_b$ is a 
scale invariant ratio of running masses. Throughout, we use the 3-loop running 
coupling normalized to $\alpha_s(M_Z)=0.1187$, matched to a 4-flavor theory 
at 4.25\,GeV. For the matching scales, we pick the default values 
$\mu_h^{\rm def}=m_b/\sqrt 2$ and 
$\mu_i^{\rm def}=\bar\mu^{\rm def}=1.5$\,GeV, which are motivated by the 
underlying dynamics of inclusive processes in the shape-function region 
\cite{Lange:2005yw,Bosch:2004th}. 

In the remainder of this section we present results for the partial decay rate 
$\Gamma_u(\Delta)$ computed by evaluating the right-hand side
of relation (\ref{eq:relation}). This is 
more informative than to focus on the value of the weight function for a  
particular choice of $P_+$. For the purpose of our discussion we use a simple 
model for the normalized photon spectrum that describes the experimental data 
reasonably well, namely
\begin{equation}\label{model}
   \frac{1}{\Gamma_s}\,\frac{d\Gamma_s}{dP_+}
   = \frac{b^b}{\Gamma(b)\Lambda^b}\,(P_+)^{b-1} 
   \exp\left( -b\,\frac{P_+}{\Lambda} \right)
\end{equation}
with $\Lambda=0.77$\,GeV and $b=2.5$. 

\subsection{Studies of the perturbative expansion}
\label{sec:pert}

\begin{figure}
\begin{center}
\epsfig{file=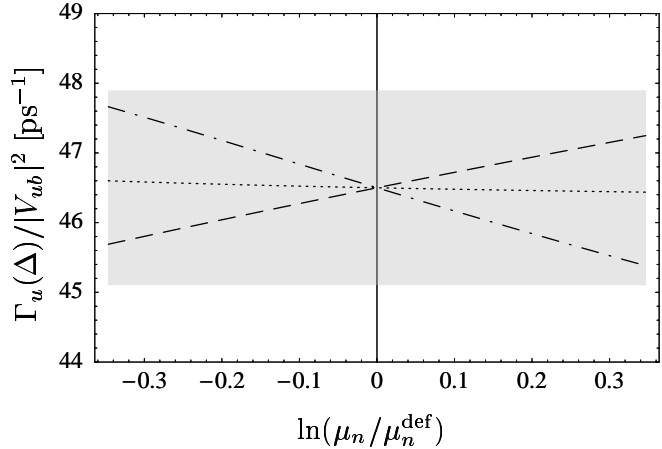,width=8.7cm}
\caption{\label{fig:scales} 
Residual scale dependence of $\Gamma_u(\Delta)$ for $\Delta=0.65$\,GeV. The 
dashed line depicts the variation of $\mu_h$ about its default value 
$m_b/\sqrt 2$, the dash-dotted line the variation of $\mu_i$ about 
$1.5$\,GeV, and the dotted line the variation of $\bar\mu$ also about 
$1.5$\,GeV. The highlighted area shows the combined perturbative uncertainty.}
\end{center}
\end{figure}

The purpose of this section is to investigate the individual
contributions to $\Gamma_u(\Delta)$ that result from the corresponding
terms in the weight function, as well as their residual dependence on
the matching scales. For $\Delta=0.65$\,GeV we find numerically
\begin{eqnarray}\label{eq:breakup}
   \frac{\Gamma_u(0.65\,{\rm GeV})}{|V_{ub}|^2\,{\rm ps}^{-1}}
   &=& 43.5\,\big( 1 + 0.158\,\hbox{\scriptsize $[\alpha_s(\mu_i)]$}  
    - 0.095\,\hbox{\scriptsize $[\alpha_s(\mu_h)]$}  
    + 0.076\,\hbox{\scriptsize $[\alpha_s^2(\mu_i)]$} \nonumber\\[-0.2cm]
   &&\hspace{1.2cm}
    \mbox{}- 0.037\,\hbox{\scriptsize $[\alpha_s(\mu_i)\alpha_s(\mu_h)]$} 
    + 0.009\,\hbox{\scriptsize [kin]}  
    - 0.043\,\hbox{\scriptsize [hadr]} \big) = 46.5 \,.
\end{eqnarray}
The terms in parenthesis correspond to the contributions to the weight 
function arising at different orders in perturbation theory and in the 
$1/m_b$ expansion, as indicated by the subscripts.
Note that the perturbative contributions from the intermediate scale are 
typically twice as large as the ones from the hard scale, which is also the
naive expectation. Indeed, the two-loop $\alpha_s^2(\mu_i)$
correction is numerically of comparable size to the one-loop
$\alpha_s(\mu_h)$ contribution. This confirms the importance of
separating the scales $\mu_i$ and $\mu_h$.
The contributions from kinematical and hadronic power corrections turn out to 
be numerically small, comparable to the two-loop corrections.

The weight function (\ref{eq:masterform}) is formally independent of the 
matching scales $\mu_h$, $\mu_i$, and $\bar\mu$. In Figure~\ref{fig:scales} we 
plot the residual scale dependence resulting from the truncation of the 
perturbative series. Each of the three scales is varied independently by a 
factor between $1/\sqrt2$ and $\sqrt2$ about its default value. The scale 
variation of $\mu_i$ is still as significant as the variation of $\mu_h$, even
though the former is known at NNLO and the latter only at NLO. We have checked 
analytically that the result
(\ref{eq:masterform}) is independent of $\mu_i$ through two-loop
order, i.e.~the residual scale dependence is an $O(\alpha_s^3(\mu_i))$ effect. 
In order to obtain a conservative estimate of the perturbative uncertainty in
our predictions we add the individual scale dependencies in quadrature. This 
gives the gray band shown in the figure.

\begin{figure}
\begin{center}
\epsfig{file=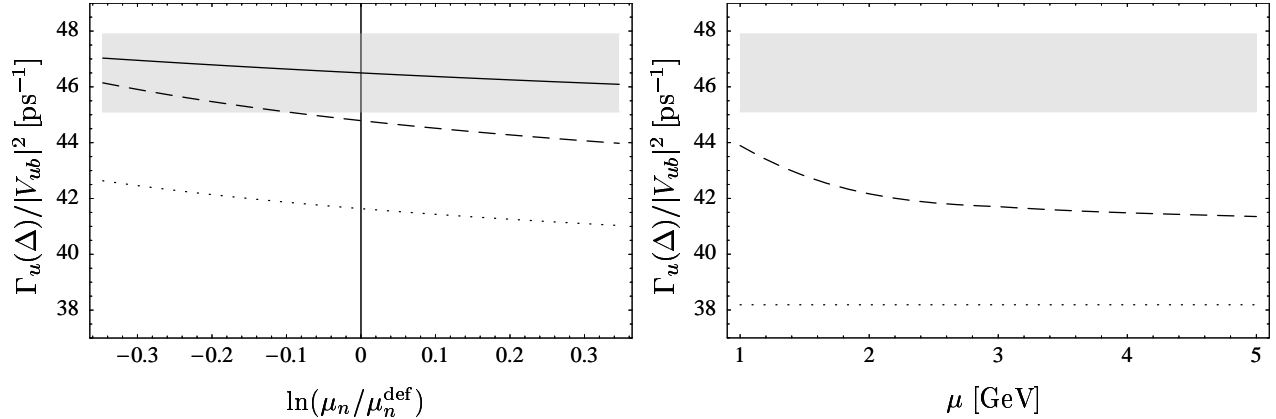,width=\textwidth}
\caption{\label{fig:onetwo} 
Convergence of the perturbative series and residual scale dependence. 
{\sc Left:} RG-improved results at LO (dotted), NLO (dashed), and NNLO (solid) 
as a function of the scales $\mu_i$, $\mu_h$, $\bar\mu$, which are varied 
simultaneously about their default values. {\sc Right:} Fixed-order results at 
tree-level (dotted) and one-loop order (dashed).}
\end{center}
\end{figure}

Figure~\ref{fig:onetwo} displays the result for $\Gamma_u(0.65\,\mbox{GeV})$ 
at different orders in RG-improved perturbation theory. At LO, we dismiss all 
$\alpha_s$ terms including the kinematical power corrections; however, leading 
logarithms are still resummed and give rise to a non-trivial dependence of 
$T_0$ on the coefficient $a$. At NLO, we include the $O(\alpha_s(\mu_h))$,
$O(\alpha_s(\mu_i))$, and $O(\alpha_s(\bar\mu))$ contributions, but drop terms 
of order $\alpha_s^2(\mu_i)$ or $\alpha_s(\mu_i)\,\alpha_s(\mu_h)$. At NNLO, 
we include all terms shown in (\ref{eq:masterform}). In studying the different 
perturbative approximations we vary the matching scales simultaneously (and 
in a correlated way) about their default values. Compared with 
Figure~\ref{fig:scales} this leads to a reduced scale variation. The gray 
bands in Figure~\ref{fig:onetwo} show the total perturbative uncertainty as
determined above. While the two-loop NNLO contributions are sizable, we 
observe a good convergence of the perturbative expansion and a reduction of 
the scale sensitivity in higher orders. 
The right-hand plot in the figure contrasts these findings with the 
corresponding results in fixed-order perturbation theory, which are obtained 
from (\ref{eq:masterform}) by setting $\mu_h=\mu_i=\bar\mu=\mu$ and truncating
the series at $O(\alpha_s(\mu))$ for consistency. We see that
the fixed-order results are also rather insensitive to the value of
$\mu$ unless this scale is chosen to be small; yet, the predicted values for
$\Gamma_u$ are significantly below those obtained in RG-improved perturbation
theory. We conclude that the small scale dependence observed in the 
fixed-order calculation does not provide a reliable estimator of the
true perturbative uncertainty. In our opinion, a fixed-order calculation at a 
high scale is not only inappropriate in terms of the underlying dynamics of 
inclusive decay processes in the shape-function region, it is also misleading 
as a basis for estimating higher-order terms in the perturbative expansion.

\subsection{Comments on the normalization of the photon spectrum}

We mentioned in the Introduction that the use of the normalized photon 
spectrum is advantageous because event fractions in $\bar B\to X_s\gamma$ 
decay can be calculated more reliably than partial decay rates. In this 
section we point out another important advantage, namely that the perturbative 
series for the weight function $W(\Delta,P_+)$ is much better behaved 
than that for $\widetilde W(\Delta,P_+)$. The difference of the two weight
functions lies in their normalizations, which are
\begin{equation}\label{hardfunx}
   W(\Delta,P_+)\propto m_b^3\, T_0(a,\mu_h) H_\Gamma(\mu_h) \,, \qquad
   \widetilde W(\Delta,P_+)\propto 
    \frac{T_0(a,\mu_h)}{[C_{7\gamma}^{\rm eff}(\mu_h)]^2\,%
    \overline{m}_b^2(\mu_h)\,|H_s(\mu_h)|^2} \,.
\end{equation}
Here $H_s$ is the hard function in the factorized expression for the structure 
function $\F_\gamma$ in (\ref{eq:gammaS}), which has been derived in 
\cite{Neubert:2004dd}.
Note that the two weight functions have a different dependence on the 
$b$-quark 
mass. In the case of $W$, three powers of $m_b$ enter through phase-space 
integrations in the total decay rate $\Gamma_s(E_*)$, and it is therefore 
appropriate to use a low-scale subtracted quark-mass definition, such as the
shape-function mass. In the case of $\widetilde W$, on the other hand, two 
powers of the running quark mass $\overline{m}_b(\mu_h)$ enter through the
definition of the dipole operator $O_{7\gamma}$, and it is appropriate to 
use a short-distance mass definition such as that provided by the 
$\overline{\rm MS}$ scheme. In practice, we write $\overline{m}_b(\mu_h)$ as 
$\overline{m}_b(m_b)$ times a perturbative series in $\alpha_s(\mu_h)$.

The most pronounced effect of the difference in normalization is that the 
weight function $\widetilde W$ receives very large radiative corrections 
at order $\alpha_s(\mu_h)$, which range between $-68\%$ and $-43\%$ when the 
scale $\mu_h$ is varied between $m_b$ and $m_b/2$. This contrasts the 
well-behaved perturbative expansion of the weight function $W$, for which the 
corresponding corrections vary between $-11\%$ and $-7\%$. In other words, 
the hard matching corrections for $\widetilde W$ are about six times larger
than those for $W$. Indeed, these corrections are so large that in our opinion
relation (\ref{oldrelation}) should not be used for phenomenological 
purposes.

The different perturbative behavior of the hard matching corrections to the 
weight functions is mostly due to the mixing of the dipole operator 
$O_{7\gamma}$ with other operators in the effective weak Hamiltonian for 
$\bar B\to X_s\gamma$ decay. In order to illustrate this fact, consider the
one-loop hard matching coefficients defined as
\begin{equation}
   W\propto 1 + k\,\frac{\alpha_s(\mu_h)}{\pi} + \dots \,, \qquad
   \widetilde W\propto 1 + \widetilde k\,\frac{\alpha_s(\mu_h)}{\pi}
    + \dots \,.
\end{equation}
With our default scale choices we have $k=-2.32+1.13=-1.19$, where the 
second contribution ($+1.13$) comes from operator mixing, which gives rise
to the terms in the last two lines in (\ref{Hgamma}). For the weight function
$W$, this contribution has the opposite sign than the other terms, so that 
the combined value of $k$ is rather small. For the weight function 
$\widetilde W$, on the other hand, we find $\widetilde k=-1.58-5.52=-7.10$. 
Here the contribution from operator mixing is dominant and has the same sign 
as the remaining terms, thus yielding a very large value of 
$\widetilde k$. Such a large $O(\alpha_s)$ correction was not 
observed in \cite{Hoang:2005pj}, because these authors chose to omit the 
contribution from operators mixing. Note that at a higher scale $\mu_h=m_b$, 
as was adopted in this reference, the situation is even worse. In that case we 
find $k=-2.81+1.19=-1.62$ and $\widetilde k=-0.66-9.13=-9.79$.

\begin{figure}
\begin{center}
\epsfig{file=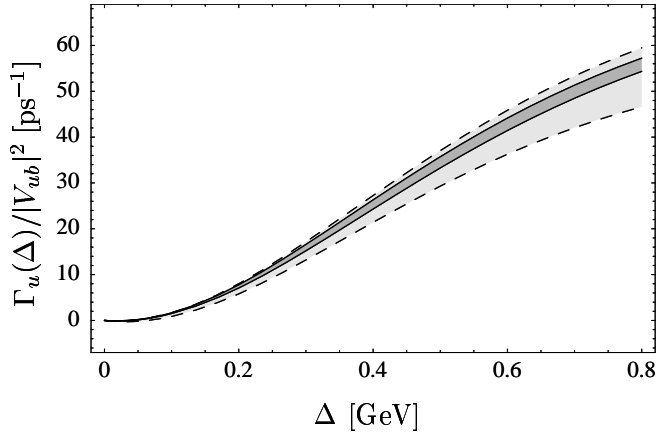,width=8.7cm}
\caption{\label{fig:band}
Perturbative uncertainties on $\Gamma_u(\Delta)$ encountered when 
using the weight function $W(\Delta,P_+)$ (dark gray) or 
$\widetilde W(\Delta,P_+)$ (light gray). In the latter case, the normalization 
of the photon spectrum is chosen such that the two predictions agree at 
$\Delta=0.65$\,GeV for central values of the matching scales.}
\end{center}
\end{figure}

A visualization of the perturbative uncertainty is depicted in
Figure~\ref{fig:band}, where predictions for $\Gamma_u(\Delta)$ are
shown using either (\ref{eq:relation}) or (\ref{oldrelation}).
In each case, the error band is obtained by 
varying the different scales about their default values, 
$\mu_n \in[\mu_n^{\rm def}/\sqrt2,\mu_n^{\rm def}\sqrt2]$, and adding the
resulting uncertainties in quadrature. The
dark-gray band bordered by solid lines denotes the perturbative
uncertainty of predictions when using the normalized photon
spectrum, as in (\ref{eq:relation}). (At the point $\Delta=0.65$\,GeV this 
uncertainty is identical to the gray band depicted in 
Figures~\ref{fig:scales} and \ref{fig:onetwo}). The light-gray band bordered
by dashed lines corresponds to the use of the absolute photon
spectrum, as in (\ref{oldrelation}). The difference in precision between the 
two methods would be even more pronounced if we used the higher default
value $\mu_h=m_b$ for the hard matching scale. Obviously, 
the use of the normalized photon spectrum will result in a more precise 
determination of $|V_{ub}|$. 

\subsection{Comments on \boldmath$\beta_0\alpha_s^2$ terms and scale 
separation\unboldmath}

The separation of different momentum scales using RG techniques, which is 
one of the key ingredients of our approach, is well motivated by the dynamics 
of charmless inclusive $B$ decays in the shape-function region. Factorized 
expressions for the $B$-decay spectra involve hard functions renormalized 
at $\mu_h$ multiplied by jet and shape functions defined at a lower scale.
While physics at or below the intermediate scale is very
similar for $\bar B\to X_u\,l^-\bar\nu$ and $\bar B\to X_s\gamma$ (as is 
manifested by the fact that the leading shape and jet functions are 
universal), the physics at the hard scale in $\bar B\to X_s\gamma$ decay is 
considerably more complicated than in semileptonic decay, and it might even 
contain effects of New Physics. Therefore it is natural to respect the
hierarchy $\mu_h\gg\mu_i$ and disentangle the various contributions, as done 
in the present work. In fact, our ability to calculate the dominant
two-loop corrections is a direct result of this scale separation.
Nevertheless, at a technical level we can reproduce the results of a
fixed-order calculation by simply setting all matching scales equal to 
a common scale, $\mu_h=\mu_i=\bar\mu=\mu$. 
In this limit, the expressions derived in this work smoothly reduce to 
those obtained in conventional perturbation theory. While factorized 
expressions for the decay rates are 
superior to fixed-order results whenever there are widely separated scales in 
the problem, they remain valid in the limit where 
the different scales become of the same order.

In a recent publication, the $O(\beta_0\alpha_s^2)$ BLM corrections 
\cite{Brodsky:1982gc} to the weight 
function $\widetilde W(\Delta,P_+)$ in (\ref{oldrelation}) were calculated in 
fixed-order perturbation theory \cite{Hoang:2005pj}. For simplicity, 
only the contribution of the operator $O_{7\gamma}$ to the 
$\bar B\to X_s\gamma$ decay rate was included in this work.
We note that without the contributions from other operators the expression for 
$\widetilde W$ is not renormalization-scale and -scheme invariant.
Neglecting operator mixing in the calculation of $\widetilde W$ is therefore 
not a theoretically consistent approximation. 
However, having calculated the exact NNLO corrections at the intermediate 
scale allows us to examine some of the 
terms proportional to $\beta_0\alpha_s^2(\mu_i)$ and compare them 
to the findings of \cite{Hoang:2005pj}. In this way
we confirm their results for the coefficients multiplying the logarithms
$\ln^n[m_b(\Delta-P_+)/\mu_i^2]$ with $n=1,2$ in (\ref{eq:masterform}). 
While the $\beta_0\alpha_s^2$ terms approximate the full two-loop coefficients 
of these logarithms arguably well, we stress that 
the two-loop constant at the intermediate 
scale is not dominated by terms proportional to $\beta_0$. Numerically we find
\begin{equation}
   c^{(2)}_0 = - 47.4 + 39.6\,\frac{\beta_0}{25/3}
   + a \left[ - 31.8 + 38.8\,\frac{\beta_0}{25/3} \right] + O(a^2) \,,
\end{equation}
which means that the approximation of keeping only the BLM terms would 
overestimate this coefficient by almost an order of magnitude and give the 
wrong sign. This shows the importance of a complete two-loop calculation, as 
performed in the present work.

We believe that the perturbative approximations adopted in our
paper, i.e.\ working to NNLO at the intermediate scale and to NLO at the hard
scale, are sufficient for practical purposes in the sense that the residual
perturbative uncertainty is smaller than other uncertainties encountered in 
the application of relation (\ref{eq:relation}). Still, one may ask what 
calculations would be required to determine the missing 
$\alpha_s^2(\mu_h)$ terms in
the normalization of the weight function in (\ref{eq:masterform}), or at 
least the terms of order $\beta_0\alpha_s^2(\mu_h)$. For the case of 
$\bar B\to X_s\gamma$ decay, the contribution of the operator $O_{7\gamma}$ 
to the normalized photon spectrum was recently calculated at two-loop order 
\cite{Melnikov:2005bx}, while the contributions from other operators are known
to $O(\beta_0\alpha_s^2)$ \cite{Ligeti:1999ea}. What is still needed are 
the two-loop corrections to the double differential (in $P_+$ and $y$) 
$\bar B\to X_u\,l^-\bar\nu$ decay rate in (\ref{eq:gammaU}).

\subsection{Subleading corrections from hadronic structures}
\label{sec:SSF}

Due to the fact that different linear combinations of the subleading shape 
functions $\hat t(\hat\omega)$, $\hat u(\hat\omega)$, and $\hat v(\hat\omega)$
enter the theoretical description of radiative and semileptonic decays
starting at order $\Lambda_{\rm QCD}/m_b$, the weight function cannot be free
of such hadronic structure functions. Consequently, we found in
(\ref{eq:masterform}) all of the above subleading shape functions,
divided by the leading shape function $\hat S(\hat\omega)$. Our default model
(\ref{defaultmodel}) for the subleading shape functions was chosen such that 
the combined effect of all hadronic power corrections could be absorbed into
a single hadronic parameter $\Lambda_{\rm SSF}^2$. More generally, 
we define a function $\delta_{\rm hadr}(\Delta)$ via (a factor 2 is inserted
for later convenience)
\begin{equation}
   \Gamma_u(\Delta) = [\Gamma_u(\Delta)]_{\rm def}
   \left[ 1 + 2\delta_{\rm hadr}(\Delta) \right] ,
\end{equation}
where $[\Gamma_u(\Delta)]_{\rm def}$ denotes the result obtained with the 
default model for the subleading shape functions.
From (\ref{eq:masterform}), one finds that
\begin{equation}\label{epsrela}
   \delta_{\rm hadr}(\Delta)
   = \frac{\int_0^\Delta\!dP_+\,(M_B-P_+)^4 \left[ 2(4-3a)\,h_t(P_+)
           + (4-a)\,h_u(P_+) + (8-13a+4a^2)\,h_v(P_+) \right]}%
          {4(1-a)(3-a)\int_0^\Delta\!dP_+\,(M_B-P_+)^5\,\hat S(P_+)} \,,
\end{equation}
where we have used that, at leading order in $\alpha_s$ and 
$\Lambda_{\rm QCD}/m_b$, the 
$\bar B\to X_s\gamma$ photon spectrum is proportional to 
$(M_B-P_+)^3\,\hat S(P_+)$. In the relation above,
$h_t(\hat\omega)\equiv\hat t(\hat\omega)-[\hat t(\hat\omega)]_{\rm def}$ etc.\ 
denote the differences between the true subleading shape functions and the 
functions adopted in our default model. By construction, these are functions
with vanishing normalization and first moment. 

The above expression for $\delta_{\rm hadr}(\Delta)$ is exact to the order we 
are working; however, in practice we do not know the precise form of the 
functions $h_i(\hat\omega)$. 
Our goal is then to find a conservative bound, 
$|\delta_{\rm hadr}(\Delta)|<\epsilon_{\rm hadr}(\Delta)$, and to interpret 
the function $\epsilon_{\rm hadr}(\Delta)$ as the relative 
hadronic uncertainty on the value of $|V_{ub}|$ extracted using relation 
(\ref{eq:relation}). To obtain the bound we scan
over a large set of realistic models for the subleading shape functions. In 
\cite{Lange:2005yw}, four different functions $h_i(\hat\omega)$ were 
suggested, which can be added or subtracted (in different combinations) to 
each of the subleading shape functions.
Together, this provides a large set of different models for these 
functions. To be conservative, we pick from this set the model 
which leads to the largest value of $|\delta_{\rm hadr}(\Delta)|$. The 
integrand in the numerator in (\ref{epsrela}) is maximized if all three 
$h_i(\hat\omega)$ functions are equal to a single function, whose choice 
depends on the value of $\Delta$. In the
denominator, we find it convenient to eliminate the shape function 
$\hat S(P_+)$ in favor of the normalized photon spectrum. Working consistently 
to leading order, we then obtain
\begin{equation}
   \epsilon_{\rm hadr}(\Delta)
   = \frac{5-5a+a^2}{(1-a)(3-a)}\,\frac{U(\mu_h,\mu_i)}{m_b^3}\,
   \max_i
   \frac{\displaystyle\left|\int_0^\Delta\!dP_+\,(M_B-P_+)^4\,h_i(P_+)\right|}%
        {\displaystyle\int_0^\Delta\!dP_+\,(M_B-P_+)^2\,
         \frac{1}{\Gamma_s(E_*)}\,\frac{d\Gamma_s}{dP_+}} \,,
\end{equation}
where as before $a=a_\Gamma(\mu_h,\mu_i)\approx 0.12$ for the default choice 
of matching scales, and $U(\mu_h,\mu_i)\approx 1.11$ \cite{Lange:2005yw}.

\begin{figure}
\begin{center}
\epsfig{file=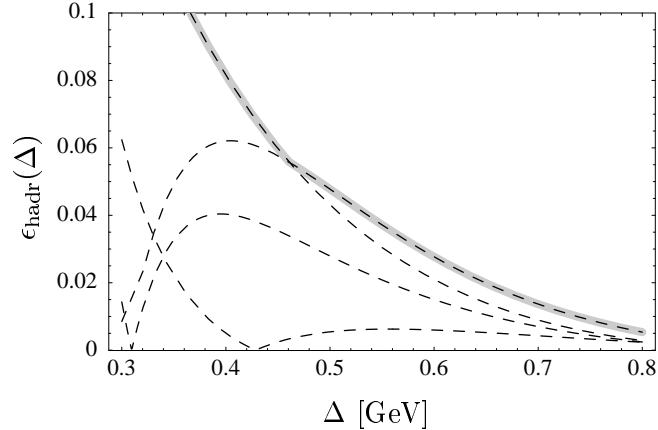,width=8.7cm}
\caption{\label{fig:sublSF}
Estimates for the hadronic uncertainty $\epsilon_{\rm hadr}(\Delta)$ obtained 
from a scan over models for the subleading shape functions. 
The dashed lines correspond to the individual results for the four 
$h_i(\hat\omega)$ functions suggested in \cite{Lange:2005yw}. The thick solid 
line, which covers one of the dashed lines, shows the maximum effect.}
\end{center}
\end{figure}

The result for the function $\epsilon_{\rm hadr}(\Delta)$ obtained this 
way is shown in Figure~\ref{fig:sublSF}. We set the matching scales to their 
default values and use the model (\ref{model}) for the photon spectrum, which 
is a good enough approximation for our purposes. 
From this estimate it is apparent that the effects of subleading shape
functions are negligible for large values $\Delta\gg\Lambda_{\rm QCD}$ and 
moderate for $\Delta\sim\Lambda_{\rm QCD}$, which is the region of interest 
for the determination of $|V_{ub}|$. In the region $\Delta<0.3$\,GeV the
accuracy of the calculation deteriorates.
For example, we find $\epsilon_{\rm hadr}=2.0\%$ for $\Delta=0.65$\,GeV, 
$\epsilon_{\rm hadr}=4.8\%$ for $\Delta=0.5$\,GeV, and
$\epsilon_{\rm hadr}=8.2\%$ for $\Delta=0.4$\,GeV.

\section{Conclusions}

Model-independent relations between weighted integrals of 
$\bar B\to X_s\gamma$ and $\bar B\to X_u\,l^-\bar\nu$ decay distributions, in 
which all reference to the leading non-perturbative shape function is avoided, 
offer one of the most promising avenues to a high-precision determination of 
the CKM matrix element $|V_{ub}|$. In order to achieve a theoretical precision 
of better than 10\%, it is necessary to include higher-order corrections in 
$\alpha_s$ and $\Lambda_{\rm QCD}/m_b$ in this approach. 

In the present work, we have calculated the weight function $W(\Delta,P_+)$ in 
the relation between the hadronic $P_+=E_X-|\vec{P}_X|$ spectra in the two 
processes, integrated over the interval $0\le P_+\le\Delta$. 
Based on QCD factorization theorems for the differential decay 
rates, we have derived an exact formula (\ref{eq:fmatch}) that allows for the 
calculation of the leading-power weight function to any order in perturbation 
theory. We have calculated the $\Delta$- and 
$P_+$-dependent terms in the weight function exactly at 
next-to-next-to-leading order (NNLO) in renormalization-group improved 
perturbation theory, including two-loop matching corrections at 
the intermediate scale $\mu_i\sim\sqrt{m_b\Lambda_{\rm QCD}}$ and three-loop 
running between the intermediate scale and the hard scale $\mu_h\sim m_b$.
The only piece missing for a complete prediction at NNLO is the two-loop
hard matching correction to the overall normalization of the weight function.
A calculation of
the $\alpha_s^2(\mu_h)$ term would require the knowledge of both decay
spectra at two-loop order, which is currently still lacking. We also include 
various sources of power corrections. Power corrections from phase-space 
factors are treated exactly. The remaining hadronic and kinematical power
corrections are given to first order in $\Lambda_{\rm QCD}/m_b$ and to the 
order in perturbation theory to which they are known.

A dedicated study of the perturbative behavior of our result for the weight 
function has been performed for the partial $\bar B\to X_u\,l^-\bar\nu$ decay 
rate $\Gamma_u(\Delta)$ as obtained from the right-hand side of relation 
(\ref{eq:relation}). It exhibits
good convergence of the expansion and reduced scale sensitivity in higher 
orders. We find that corrections of order $\alpha_s^2(\mu_i)$ at the 
intermediate scale are typically as important as first-order $\alpha_s(\mu_h)$ 
corrections at the hard scale. We have also seen that fixed-order 
perturbation theory significantly underestimates the value of 
$\Gamma_u(\Delta)$, even though the apparent stability with respect to scale
variations would suggest a good perturbative convergence. In order to obtain 
a well-behaved expansion in powers of $\alpha_s(\mu_h)$, it is important to 
use the normalized photon spectrum in relation (\ref{eq:relation}). A similar 
relation involving the differential $\bar B\to X_s\gamma$ decay rate
receives uncontrollably large matching 
corrections at the hard scale and is thus not suitable for phenomenological
applications.
At next-to-leading order in the $1/m_b$ expansion, the weight function 
receives terms involving non-perturbative subleading shape functions, which 
cannot be eliminated. Our current ignorance about the functional forms of 
these functions leads to a hadronic uncertainty, which we have 
estimated by scanning over a large set of models. We believe that a 
reasonable estimate of the corresponding relative uncertainty 
$\epsilon_{\rm hadr}$ on $|V_{ub}|$ is given by the solid line in 
Figure~\ref{fig:sublSF}.

Let us summarize our main result for the partial $\bar B\to X_u\,l^-\bar\nu$ 
decay rate with a cut $P_+\le 0.65$\,GeV, which is close to the charm 
threshold $M_D^2/M_B$, and present a detailed list of the various sources of 
theoretical uncertainties. We find 
\begin{eqnarray}
   \Gamma_u(0.65\,{\rm GeV})
   &=& \left( 46.5\pm 1.4\,\hbox{\scriptsize [pert]}\,
    \pm 1.8\, \hbox{\scriptsize [hadr]}\,
    \pm 1.8\, \hbox{\scriptsize [$m_b$]}\,
    \pm 0.8\, \hbox{\scriptsize [pars]}\,
    \pm 2.8\, \hbox{\scriptsize [norm]}\,
    \right) |V_{ub}|^2\,{\rm ps}^{-1} \nonumber\\
   &=& (46.5\pm 4.1)\,|V_{ub}|^2\,{\rm ps}^{-1}\,,
\end{eqnarray}
where the central value is derived assuming that the $\bar B\to X_s\gamma$ 
photon spectrum can be accurately described by the function (\ref{model}). 
The errors refer to the perturbative uncertainty as estimated in 
Section~\ref{sec:pert}, the uncertainty due to the ignorance about 
subleading shape functions as discussed in Section~\ref{sec:SSF}, the error 
in the value of the $b$-quark mass, 
other parametric uncertainties from variations of $m_c$, $m_s$, and 
$\lambda_1$, and finally a 6\% uncertainty in the calculation of the 
normalization of the photon spectrum \cite{Neubert:2004dd}. To a good
approximation the errors scale with the central value. The above numbers 
translate into a combined theoretical uncertainty of 4.4\% on $|V_{ub}|$ when 
added in quadrature. 

\vspace{0.5cm}
\noindent 
{\em Acknowledgments:} 
We thank the Institute of Nuclear Theory at the University of Washington, 
where part of this research has been performed. The work of M.N.\ was 
supported in part by a Research Award of the Alexander von Humboldt Foundation.
The work of B.O.L.\ was supported in part by funds provided by the 
U.S.~Department of Energy under cooperative research agreement 
DE-FC02-94ER40818. The research of M.N.\ and G.P.\ was supported by the 
National Science Foundation under Grant PHY-0355005.


\begin{thebibliography}{99}

\bibitem{Neubert:1993ch}
M. Neubert,
Phys.\ Rev.\ D {\bf 49}, 3392 (1994)
[hep-ph/9311325].

\bibitem{Neubert:1993um}
M.~Neubert,
Phys.\ Rev.\ D {\bf 49}, 4623 (1994)
[hep-ph/9312311].

\bibitem{Bigi:1993ex}
I.~I.~Y.~Bigi, M.~A.~Shifman, N.~G.~Uraltsev and A.~I.~Vainshtein,
Int.\ J.\ Mod.\ Phys.\ A {\bf 9}, 2467 (1994)
[hep-ph/9312359].

\bibitem{Mannel:1999gs}
T.~Mannel and S.~Recksiegel,
Phys.\ Rev.\ D {\bf 60}, 114040 (1999)
[hep-ph/9904475].

\bibitem{Aglietti:2002md}
U.~Aglietti, M.~Ciuchini and P.~Gambino,
Nucl.\ Phys.\ B {\bf 637}, 427 (2002)
[hep-ph/0204140].

\bibitem{Bosch:2004bt}
S.~W.~Bosch, B.~O.~Lange, M.~Neubert and G.~Paz,
Phys.\ Rev.\ Lett.\  {\bf 93}, 221801 (2004)
[hep-ph/0403223].

\bibitem{Lange:2005yw}
B.~O.~Lange, M.~Neubert and G.~Paz,
hep-ph/0504071, Phys.\ Rev.~D (in press).

\bibitem{Leibovich:1999xf}
A.~K.~Leibovich, I.~Low and I.~Z.~Rothstein,
Phys.\ Rev.\ D {\bf 61}, 053006 (2000)
[hep-ph/9909404].

\bibitem{Leibovich:2000ey}
A.~K.~Leibovich, I.~Low and I.~Z.~Rothstein,
Phys.\ Lett.\ B {\bf 486}, 86 (2000)
[hep-ph/0005124]. 

\bibitem{Neubert:2001sk}
M.~Neubert,
Phys.\ Lett.\ B {\bf 513}, 88 (2001)
[hep-ph/0104280].

\bibitem{Hoang:2005pj}
A.~H.~Hoang, Z.~Ligeti and M.~Luke,
Phys.\ Rev.\ D {\bf 71}, 093007 (2005)
[hep-ph/0502134].

\bibitem{Kagan:1998ym}
A.~L.~Kagan and M.~Neubert,
Eur.\ Phys.\ J.\ C {\bf 7}, 5 (1999)
[hep-ph/9805303].

\bibitem{Neubert:2004dd}
M.~Neubert,
Eur.\ Phys.\ J.\ C {\bf 40}, 165 (2005)
[hep-ph/0408179].

\bibitem{Bauer:2003pi}
C.~W.~Bauer and A.~V.~Manohar,
Phys.\ Rev.\ D {\bf 70}, 034024 (2004)
[hep-ph/0312109].

\bibitem{Bosch:2004th}
S.~W.~Bosch, B.~O.~Lange, M.~Neubert and G.~Paz,
Nucl.\ Phys.\ B {\bf 699}, 335 (2004)
[hep-ph/0402094].

\bibitem{Neubert:2005nt}
M.~Neubert,
hep-ph/0506245, Phys.\ Rev.\ D (in press).

\bibitem{Neubert:2004sp}
M.~Neubert,
Phys.\ Lett.\ B {\bf 612}, 13 (2005)
[hep-ph/0412241].

\bibitem{Korchemsky:wg}
G.~P.~Korchemsky and A.~V.~Radyushkin,
Nucl.\ Phys.\ B {\bf 283}, 342 (1987).

\bibitem{Korchemskaya:1992je}
I.~A.~Korchemskaya and G.~P.~Korchemsky,
Phys.\ Lett.\ B {\bf 287}, 169 (1992).

\bibitem{Moch:2004pa}
S.~Moch, J.~A.~M.~Vermaseren and A.~Vogt,
Nucl.\ Phys.\ B {\bf 688}, 101 (2004)
[hep-ph/0403192].

\bibitem{DeFazio:1999sv}
F.~De Fazio and M.~Neubert,
JHEP {\bf 9906}, 017 (1999)
[hep-ph/9905351].

\bibitem{Bauer:2001yt}
C.~W.~Bauer, D.~Pirjol and I.~W.~Stewart,
Phys.\ Rev.\ D {\bf 65}, 054022 (2002)
[hep-ph/0109045].

\bibitem{Bauer:2001mh}
C.~W.~Bauer, M.~E.~Luke and T.~Mannel,
Phys.\ Rev.\ D {\bf 68}, 094001 (2003)
[hep-ph/0102089].

\bibitem{Leibovich:2002ys}
A.~K.~Leibovich, Z.~Ligeti and M.~B.~Wise,
Phys.\ Lett.\ B {\bf 539}, 242 (2002)
[hep-ph/0205148].

\bibitem{Bauer:2002yu}
C.~W.~Bauer, M.~Luke and T.~Mannel,
Phys.\ Lett.\ B {\bf 543}, 261 (2002)
[hep-ph/0205150].

\bibitem{Neubert:2002yx}
M.~Neubert,
Phys.\ Lett.\ B {\bf 543}, 269 (2002)
[hep-ph/0207002].

\bibitem{Burrell:2003cf}
C.~N.~Burrell, M.~E.~Luke and A.~R.~Williamson,
Phys.\ Rev.\ D {\bf 69}, 074015 (2004)
[hep-ph/0312366].

\bibitem{Lee:2004ja}
K.~S.~M.~Lee and I.~W.~Stewart,
Nucl.\ Phys.\ B {\bf 721}, 325 (2005)
[hep-ph/0409045].

\bibitem{Bosch:2004cb}
S.~W.~Bosch, M.~Neubert and G.~Paz,
JHEP {\bf 0411}, 073 (2004)
[hep-ph/0409115].

\bibitem{Beneke:2004in}
M.~Beneke, F.~Campanario, T.~Mannel and B.~D.~Pecjak,
JHEP {\bf 0506}, 071 (2005)
[hep-ph/0411395].

\bibitem{Chay:2005ck}
J.~Chay, C.~Kim and A.~K.~Leibovich,
Phys.\ Rev.\ D {\bf 72}, 014010 (2005)
[hep-ph/0505030].

\bibitem{Falk:1992wt}
A.~F.~Falk and M.~Neubert,
Phys.\ Rev.\ D {\bf 47}, 2965 (1993)
[hep-ph/9209268].

\bibitem{Gamiz:2004ar}
E.~Gamiz, M.~Jamin, A.~Pich, J.~Prades and F.~Schwab,
Phys.\ Rev.\ Lett.\  {\bf 94}, 011803 (2005)
[hep-ph/0408044].

\bibitem{Aubin:2004ck}
C.~Aubin {\it et al.}  [HPQCD Collaboration],
Phys.\ Rev.\ D {\bf 70}, 031504 (2004)
[hep-lat/0405022].

\bibitem{Brodsky:1982gc}
S.~J.~Brodsky, G.~P.~Lepage and P.~B.~Mackenzie,
Phys.\ Rev.\ D {\bf 28}, 228 (1983).

\bibitem{Melnikov:2005bx}
K.~Melnikov and A.~Mitov,
Phys.\ Lett.\ B {\bf 620}, 69 (2005)
[hep-ph/0505097].

\bibitem{Ligeti:1999ea}
Z.~Ligeti, M.~E.~Luke, A.~V.~Manohar and M.~B.~Wise,
Phys.\ Rev.\ D {\bf 60}, 034019 (1999)
[hep-ph/9903305].

\end{thebibliography}
\end{document}